\documentstyle[12pt,epsfig]{article}
\def\thebibliography#1{\leftline{\large\it References}\list
  {[\arabic{enumi}]}{\settowidth\labelwidth{[#1]}\leftmargin\labelwidth
    \advance\leftmargin\labelsep
    \usecounter{enumi}}
    \def\newblock{\hskip .11em plus .33em minus .07em}
    \sloppy\clubpenalty4000\widowpenalty4000}

\newcommand{\be}{\begin{eqnarray}}
\newcommand{\ba}{\begin{array}}
\newcommand{\ea}{\end{array}}
\newcommand{\ee}{\end{eqnarray}}

\newcommand{\dslash}{\partial \hskip -0.5em /}

\newcommand{\Tr}{{\rm Tr}}

\newcommand{\La}{{\cal L}}

\newcommand{\bbox}[1]{\mbox{\boldmath$#1$\unboldmath}}

\begin{document}
\rightline{May 1998}
\rightline{UNITU-THEP-10/1998}
\rightline{hep-ph/9805251}
\vskip 1.0truecm
\centerline{\Large\bf Strangeness Contribution to the}
\vskip0.5cm
\centerline{\Large\bf
Polarized Nucleon Structure Function 
$\mbox{\boldmath $g_1(x)$}$~$^{\textstyle\dagger}$}
\baselineskip=16 true pt
\vskip 1.0cm
\centerline{O.\ Schr\"oder, H.\ Reinhardt, and H.\ Weigel}
\vskip 0.5cm
\centerline{Institute for Theoretical Physics}
\centerline{T\"ubingen University}
\centerline{Auf der Morgenstelle 14}
\centerline{D-72076 T\"ubingen, Germany}
\vskip 1.0cm
\baselineskip=16pt
\centerline{\bf ABSTRACT}
\vskip 0.5cm
\parbox[t]{14cm}{
\baselineskip=20pt
The three flavor version of the Nambu--Jona--Lasinio chiral soliton 
model for baryons is employed to calculate the twist--2 contribution
to the polarized nucleon structure function $g_1(x)$. In particular the 
role of the strange quark degree of freedom as a collective excitation 
of the chiral soliton is investigated in the context of flavor 
symmetry breaking. The model prediction for
$g_1(x)$ refers to a low momentum scale $Q_0^2$. The leading order
corrections to the scale dependence is computed along the QCD evolution 
program allowing to compare with data from SLAC.
}
\vskip 2.0cm
\leftline{\it PACS: 11.30.Cp, 12.39.Ki.}

\vfill
\noindent
--------------

\noindent
$^{\textstyle\dagger}$
{\footnotesize{Supported by the Deutsche Forschungsgemeinschaft
(DFG) under contract Re 856/2-3.}}
\eject

\baselineskip=16pt

\leftline{\large\it 1. Introduction}
\medskip

The investigation of the strange quark contribution to the polarized 
structure function is mainly motivated by the empirical results obtained 
in the context with the {\it proton spin puzzle}, {\it cf.} ref
\cite{El96} for a review. In this context one considers nucleon matrix 
elements of axial currents, which are obtained as the zeroth moment of 
polarized structure functions. The {\it puzzle} firstly refers to the 
smallness of the observed nucleon matrix element of the axial singlet 
current. Already early studies \cite{Br88} in the Skyrme model (the 
simplest version of a chiral soliton model) indicated that chiral soliton 
models are capable of reproducing that result. The data analysis secondly 
revealed that the strange quark might make a sizable contribution to the 
axial singlet charge of the nucleon; up to a third of that of the down 
quark. This surprising result was obtained using SU(3) symmetric baryon 
wave--functions; the inclusion of symmetry breaking effects reduces this 
ratio. Certainly it is interesting to compute the full dependence of the 
corresponding structure functions on the Bjorken variable. At this point 
the collective approach to incorporate strange degrees of freedom in chiral 
soliton models \cite{Gu84} becomes very attractive\footnote{{\it Cf.} ref 
\cite{We96} for a recent review on soliton models in flavor SU(3).}. It 
not only allows one to account for such symmetry breaking effects in the 
nucleon wave--function \cite{Ya88} but in particular to make explicit the 
strange quark contribution to nucleon structure functions. This is a major 
advantage over other low energy models for the nucleon like {\it e.g.} the 
MIT bag model \cite{Ch74,Ja75}. 

The polarized structure function $g_1$ is extracted from
the hadronic tensor for electron--nucleon scattering,
\be
W_{\mu\nu}(q)=\frac{1}{4\pi}\int d^4 \xi \
{\rm e}^{iq\cdot\xi}
\langle N |\left[J_\mu(\xi),J^{\dag}_\nu(0)\right]|N\rangle\ ,
\label{deften}
\ee
where $J_\mu={\bar q}(\xi)\gamma_\mu {\cal Q} q(\xi)$ is the
electromagnetic current with
${\cal Q}={\rm diag}\left(\frac{2}{3},-\frac{1}{3},-\frac{1}{3}\right)$ 
being the quark charge matrix. The nucleon state, denoted by $|N\rangle$ 
in eq (\ref{deften}), is characterized by its momentum ($P_\mu$) and spin
($S_\mu$). The polarized structure function $g_1$ parameterizes the 
longitudinal part of the antisymmetric piece, 
\be
W^{(A)}_{\mu\nu}(q)=(W_{\mu\nu}-W_{\nu\mu})/2i=
i\epsilon_{\mu\nu\lambda\sigma}\frac{q^{\lambda}M_N}{P\cdot q}
\left\{g_1(x,Q^2)S^{\sigma}+{\rm transverse~part}\right\} .
\label{had}
\ee
Here $q_\mu$ refers to the momentum transferred to the nucleon by the 
virtual photon with $Q^2=-q^2$. The Bjorken variable is defined 
as $x=Q^2/P\cdot q$. We are interested in the leading twist 
contribution to $g_1$. It is extracted from the hadronic tensor by 
assuming the Bjorken limit which corresponds to the kinematic regime
\be
q_0=|\mbox{\boldmath $q$}| - M_N x
\quad {\rm with}\quad
|\mbox{\boldmath $q$}|\rightarrow \infty 
\quad {\rm and}\ \ x\ \ {\rm fixed} \ .
\label{bjlimit}
\ee
In this limit eq (\ref{had}) is straightforwardly inverted yielding
\be
g_1(x)=\frac{M_N}{2}\, \epsilon^{\mu\nu30}\, W^{(A)}_{\mu\nu}(q)
\label{g1wa}
\ee
for the kinematical conditions
$q=(q^0,0,0,|\bbox{q}|)$, $P=(M_N,0,0,0)$ and 
$S=(0,0,0,1)$, {\it i.e.} the nucleon rest frame with the nucleon spin and 
the transferred momentum being aligned, hence the notion longitudinal.

In chiral soliton models strange quark effects enter the structure 
functions in two ways, firstly there are direct contributions to the 
currents $J_\mu$ which do not exist in the two flavor model and 
secondly the structure of the nucleon state $|N\rangle$ changes when 
generalizing from flavor SU(2) to SU(3). For the present study we will 
consider the Nambu--Jona--Lasinio (NJL) model \cite{Na61} in its 
bosonized form \cite{Eb86} which is well known to possess a chiral 
soliton solution \cite{Re88}. For comprehensive lists of references we 
refer to recent reviews \cite{Al96,Ch96}. In this model the defining 
Lagrangian contains only quark fields hence all quantities can formally be 
expressed in terms of these fields. In particular the formal expression 
for the current is that of a free Dirac theory which makes the commutator 
in eq (\ref{deften}) feasible. It is this feature which actually allows 
us to compute structure functions from a chiral soliton. In other soliton 
models, which do not possess such a clear connection to the quark flavor 
dynamics, the computation of structure functions seems infeasible due to 
the complicated structure of the current operator. In the two flavor 
version of the model various nucleon structure functions have been 
discussed \cite{We96a,Di96,We97,Po96,Wa98,Ga98} while the present 
study represents the first attempt to consider the effects of strange 
quarks.

The extension to flavor SU(3) is particularly interesting because the 
calculation \cite{We97} of $g_1$ in the two flavor model already 
yielded reasonable agreement with experiment. On the other hand one
expects strange degrees of freedom to have non--negligible impact 
on nucleon properties. 

\bigskip
\leftline{\large\it 2. The NJL Chiral Soliton}
\medskip

In the three dimensional flavor space with {\it up, down} and {\it strange}
quarks the NJL model Lagrangian reads
\be
\La = \bar q\, (i\dslash - m^0)\, q +
      2\, G_{\rm NJL}\, \sum _{i=0}^{8}
\left( (\bar q \frac {\lambda^i}{2} q )^2
      +(\bar q \frac {\lambda^i}{2} i\gamma _5 q )^2 \right) .
\label{NJL}
\ee
Here $q$, $\hat m^0={\rm diag}(m_u^0,m_d^0,m_s^0)$ and $G_{\rm NJL}$
denote the quark field, the current quark mass matrix and a dimensionful
coupling constant, respectively. In what follows we will assume the
isospin limit $m_u^0=m_d^0$. Of course, we will consider flavor
symmetry breaking by allowing the strange current quark mass
to be different, {\it i.e.} $m_s^0\ne m_u^0$.

Using functional integration techniques we obtain the bosonized 
version of the NJL model action \cite{Eb86}
\be
{\cal A}[{\cal M}]&=&\Tr_\Lambda\log(iD)+\frac{1}{4G_{\rm NJL}}
\int d^4x\ {\rm tr}
\left\{m^0\left({\cal M}+{\cal M}^{\dag}\right)
-{\cal M}{\cal M}^{\dag}-(m^0)^2\right\}\ ,
\label{bosact} \\
D&=&i\dslash-\left({\cal M}+{\cal M}^{\dag}\right)
-\gamma_5\left({\cal M}-{\cal M}^{\dag}\right)\ .
\label{dirac}
\ee
The composite scalar (${\cal S}$) and pseudoscalar (${\cal P}$) meson 
fields are contained in ${\cal M}={\cal S} +i{\cal P}$, and appear as 
quark--antiquark bound states. Apparently ${\cal M}$ represents a 
$3\times3$ matrix field in flavor space which behaves as the sum of 
scalar (${\cal S}$) and pseudoscalar (${\cal P}$) quark bilinears 
under chiral transformations. For regularization, which is indicated by 
the cut--off $\Lambda$, we will adopt the proper--time scheme \cite{Sch51}. 
The free parameters of the model are the current quark mass matrix $m^0$, the 
coupling constant $G_{\rm NJL}$ and the cut--off $\Lambda$. The equation 
of motion for the scalar field ${\cal S}$ may be considered as the 
gap--equation for the order parameter $\langle {\bar q} q\rangle$ of 
chiral symmetry breaking. This equation relates the vacuum expectation 
value, $\langle {\cal M}\rangle=M={\rm diag}(M_u,M_d=M_u,M_s)$ to the model 
parameters ${\hat m}^0$, $G_{\rm NJL}$ and $\Lambda$. For apparent reasons 
$M$ is called the {\em constituent} quark mass matrix. The occurrence of 
this vacuum expectation value reflects the spontaneous breaking of chiral 
symmetry and causes the pseudoscalar fields to emerge as (would--be) 
Goldstone bosons. At this stage we expand ${\cal A}$ to quadratic order in 
${\cal P}$ around $M$. Then the model parameters are related to physical 
quantities like the pion mass, $m_\pi=135{\rm MeV}$ and the pion decay 
constant, $f_\pi=93{\rm MeV}$. In account of the gap--equation this leaves 
one undetermined parameter which we choose to be the up constituent quark 
mass $M_u$ \cite{Eb86}. Typical values are $M_u=350\sim450{\rm MeV}$.
The kaon mass $m_K=495{\rm MeV}$ allows us to fix the strange current 
quark mass which, subject to the gap--equation in the strange sector, 
also determines the strange constituent quark mass $M_s$. Using the 
typical values for $M_u$ the model underestimates the kaon decay constant 
$f_K=114{\rm MeV}$ by about 10--15\% \cite{We92}. 

Turning to the baryon sector of the model we adopt the hedgehog 
{\it ansatz} for the meson fields ${\cal M}=\xi M \xi$ 
\be
\xi=\xi_H(\bbox{r})= {\rm exp}\left(\frac{i}{2}\, \bbox{\tau}
\cdot{\hat{\bbox{r}}}\, \Theta(r)\right) 
\label{hedgehog}
\ee
in order to determine the chiral soliton. For static meson configurations
as (\ref{hedgehog}) it is straightforward to deduce the classical energy 
$E[\Theta]$ functional associated with the action (\ref{bosact}) 
\cite{Re89}
\be
E[\Theta]&=&
\frac{N_C}{2}\epsilon_{\rm v}
\Big(1+{\rm sgn}(\epsilon_{\rm v})\Big)
+\frac{N_C}{2}\int^\infty_{1/\Lambda^2}
\frac{ds}{\sqrt{4\pi s^3}}\sum_\nu{\rm exp}
\left(-s\epsilon_\nu^2\right)
\nonumber \\* && \hspace{3.5cm}
+\ m_\pi^2 f_\pi^2\int d^3r  \Big(1-{\rm cos}\, \Theta(r)\Big) .
\label{efunct}
\ee
Here $\epsilon_\mu$ refer to the energy eigenvalues of the 
Dirac Hamiltonian 
\be
h_0=\mbox{\boldmath $\alpha$}\cdot\mbox{\boldmath $p$}+M_u\, \beta\,
{\rm exp}\Big(i\gamma_5\mbox{\boldmath $\tau$}
\cdot{\hat{\mbox{\boldmath $r$}}}\, \Theta(r)\Big)\, \hat{T}
+M_s\, \beta\, \hat{S}\,
\label{hamil}
\ee
which is derived from the operator (\ref{dirac}) 
$D=\beta(i\partial_t-h_0)$ upon substituting the hedgehog {\it ansatz} 
(\ref{hedgehog}). From the appearance of the strange and non--strange 
projectors $\hat{S}={\rm diag}_{\rm fl}(0,0,1)$ and
$\hat{T}={\rm diag}_{\rm fl}(1,1,0)$ we observe that the strange quarks 
are not effected by the hedgehog field. In eq (\ref{efunct})
the subscript ``${\rm v}$" denotes the valence quark level. This state
is the distinct level bound in the soliton background, {\it i.e.}
$-m<\epsilon_{\rm v}<m$. Similar to the energy functional (\ref{efunct})
other quantities also separate into contributions associated with the 
explicit occupation of the valence level and a (regularized) piece due
to the vacuum being polarized by the meson fields. The chiral soliton, 
$\Theta(r)$, is finally obtained by self--consistently extremizing 
$E[\Theta]$ \cite{Re88}.

\bigskip
\leftline{\large\it 3. The Nucleon State in Flavor SU(3)}
\medskip

States with nucleon quantum numbers are generated from the static 
configuration (\ref{hedgehog}) by introducing collective coordinates 
for the large amplitude fluctuations of the soliton \cite{Ad83,Gu84}. 
This approach may be viewed as an approximation to the unknown time 
dependent solution to the equations of motion for the meson fields. 
Subsequently these coordinates are treated quantum--mechanically. The 
large amplitude fluctuations are associated with the rotations in flavor 
space because the spatial rotations can be absorbed into the former as 
a consequence of the hedgehog {\it ansatz} (\ref{hedgehog}). For the 
two--flavor NJL chiral soliton model this quantization approach has
been performed in ref \cite{Re89}. In the more involved case of three 
flavors we have \cite{We92}
\be
\xi(\bbox{r},t)=A(t)\, \xi_H(\bbox{r})\, A^\dagger(t)\ ,
\label{colcor}
\ee
where $A(t)$ is a $3\times3$ matrix in flavor space. Substituting
this configuration and transforming to the flavor rotating 
frame $q^\prime =A(t)q$ reveals that the eigenvalues of the 
modified Dirac Hamiltonian
\be
h^\prime=h_0+h_{\rm rot}+h_{\rm SB}
\label{frf1}
\ee
with 
\be
h_{\rm rot}&=&-iA(t)\frac{d}{dt}A^\dagger(t)=
\frac{1}{2}\sum_{a=1}^8\lambda_a\Omega^a
\label{frf2} \\
h_{\rm SB}&=&\frac{1}{\sqrt{3}}\Big(M_u-M_s\Big)
{\cal T}\beta\left\{\sum_{i=1}^3D_{8i}\lambda_i
+\sum_{\alpha=4}^7D_{8\alpha}\lambda_\alpha
+\Big(D_{88}-1\Big)\lambda_8\right\}{\cal T}^\dagger
\label{frf3}
\ee
enter the functional trace 
(\ref{bosact}) \cite{We92}. Besides the angular velocities $\Omega^a$ 
also the adjoint representation of the collective coordinates 
$D_{ab}=(1/2){\rm tr}(\lambda_a A \lambda_b A^\dagger)$ have been
used to simplify the additional parts of the Dirac Hamiltonian.
For convenience the chiral rotation 
${\cal T}=(\xi_H^\dagger+\xi_H)/2
+\gamma_5(\xi_H^\dagger-\xi_H)/2$ has been introduced in 
eq (\ref{frf3}).
Using these definitions it is straightforward to extract the Lagrangian 
$L(A,\Omega^a)$ for the collective coordinates from the action functional 
(\ref{bosact}). As the rotations (\ref{colcor}) are assumed to proceed 
adiabatically the expansion of the action functional is terminated at 
quadratic order in $\Omega^a$. In addition the functional trace in 
(\ref{bosact}) is expanded in the difference $M_s-M_u$ which measures 
the flavor symmetry breaking originating from different constituent quark 
masses\footnote{The flavor symmetry breaking stemming from different 
current quark masses resides in the local piece in (\ref{bosact}).}.
Essentially this represents an expansion in $h_{\rm rot}+h_{\rm SB}$.

By Legendre transformation to the SU(3) right generators,
$R_b=-\partial L(A,\Omega^a)/\partial \Omega^b$
the Hamilton operator in the space of the collective coordinates is 
deduced. It has the form \cite{We96}
\be
H(A,R_a)&=&E
+\frac{1}{2}\left[\frac{1}{\alpha^2}-\frac{1}{\beta^2}\right]
\sum_{i=1}^3 R_i^2
+\frac{1}{2\beta^2}\sum_{a=1}^8 R_a^2
-\frac{3}{8\beta^2}
\nonumber \\ &&
+\frac{\alpha_1}{2\alpha^2}\sum_{i=1}^3D_{8i}
\left(2R_i+\alpha_1D_{8i}\right)
+\frac{\beta_1}{2\beta^2}\sum_{\alpha=4}^7D_{8\alpha}
\left(2R_\alpha+\beta_1D_{8\alpha}\right)
+\frac{1}{2}\gamma \left(1-D_{88}\right)
\nonumber \\ &&
+\frac{1}{2}\gamma_S\left(1-D_{88}^2\right)
+\frac{1}{2}\gamma_T\sum_{i=1}^3D_{8i}D_{8i}
+\frac{1}{2}\gamma_{TS}\sum_{\alpha=4}^7D_{8\alpha}D_{8\alpha}
\label{collham}
\ee
together with the constraint $R_8=\sqrt3/2$ for $B=1$ and $N_C=3$. 
This constraint restricts the allowed states to those with half--integer 
spin. The quantities $\alpha^2,\ldots,\gamma_{TS}$ are functionals of the 
self--consistent chiral angle. For details of their evaluation and 
numerical results we refer to ref \cite{We92}. Here it is only important 
to note that for the self--consistent soliton these constants of 
proportionality are dominated by their valence quark contributions.

The most important feature of the collective Hamiltonian is that 
it can be diagonalized exactly yielding as eigenstates the low--lying 
$\frac{1}{2}^+$ and $\frac{3}{2}^+$ baryons. Due to the presence of
flavor symmetry breaking these are no longer pure octet (decouplet) 
states but acquire sizable admixture of states with identical 
quantum numbers in the higher dimensional SU(3) representations 
like ${\overline{\bbox{10}}}$ and $\bbox{27}$. This diagonalization 
is a generalization of the Yabu--Ando approach \cite{Ya88} and 
is comprehensively described in refs \cite{We92,We96}. All nucleon 
matrix elements to be computed henceforth will employ these exact 
eigenstates.

\bigskip
\leftline{\large\it 4. Valence Quark Approximation to $g_1(x)$}
\medskip

In order to compute $g_1$ in the present model we have to bear in 
mind that we are dealing with localized field configurations. The 
resulting expression for the hadronic tensor in the Bjorken limit has 
been obtained in \cite{Ja75}. Its antisymmetric component becomes 
\cite{Ja75}
\be
W^{(A)}_{\mu\nu}(q)&=&\int \frac{d^4k}{(2\pi)^4} \
\epsilon_{\mu\rho\nu\sigma}\ k^\rho\
{\rm sgn}\left(k_0\right) \ \delta\left(k^2\right)
\int_{-\infty}^{+\infty} dt \ {\rm e}^{i(k_0+q_0)t}
\nonumber \\* &&
\times \int d^3x_1 \int d^3x_2 \
{\rm exp}\left[-i(\mbox{\boldmath $k$}+\mbox{\boldmath $q$})\cdot
(\mbox{\boldmath $x$}_1-\mbox{\boldmath $x$}_2)\right]
\nonumber \\* &&
\times \langle N |\left\{
{\bar \Psi}(\mbox{\boldmath $x$}_1,t){\cal Q}^2\gamma^\sigma\gamma^{5}
\Psi(\mbox{\boldmath $x$}_2,0)+
{\bar \Psi}(\mbox{\boldmath $x$}_2,0){\cal Q}^2\gamma^\sigma\gamma^{5}
\Psi(\mbox{\boldmath $x$}_1,t)\right\}| N \rangle \ ,
\label{stpnt}
\ee
where $\epsilon_{\mu\rho\nu\sigma}\gamma^\sigma \gamma^5$
is the
antisymmetric combination of $\gamma_\mu\gamma_\rho\gamma_\nu$. The 
expression (\ref{stpnt}) is essentially obtained by applying Wick's 
theorem to the current commutator in eq (\ref{deften}). For the intermediate
quark the free correlator is substituted because in the Bjorken limit
this quark is far off--shell and hence not sensitive to momenta which 
are of the scale as those attributed to the soliton. Finally a collective 
coordinate describing the position of the soliton in coordinate space is 
introduced and integrated over. 

We have already noted that the constants of proportionality 
entering the collective Hamiltonian (\ref{collham}) are dominated
by their valence quark contribution. This dominance is even more 
pronounced for axial matrix elements like
$\langle N |{\bar q}\gamma_5\gamma_3\lambda_a q| N\rangle$ 
\cite{Al96,Ch96,Bl96}. For these axial current matrix elements the 
vacuum contribution is commonly found to be 10\% of the total;
for some flavor combinations even less. This establishes the assumption
that the vacuum contribution to the polarized structure functions is
negligible and manifests itself in the valence quark approximation for 
the structure function $g_1(x)$. This approximation is defined by 
substituting the valence quark wave--function
\be
\Psi_{\rm v}(\bbox{r},t)={\rm e}^{-i\epsilon_{\rm v}t}A(t)
\Phi_{\rm v}(\bbox{r}) \quad {\rm with} \quad
\Phi_{\rm v}(\bbox{r})=\psi_{\rm v}(\bbox{r})
+\sum_{\mu\ne{\rm v}}\psi_\mu(\bbox{r})\, 
\frac{\langle\mu|h_{\rm rot}+h_{\rm SB}|{\rm v}\rangle}
{\epsilon_{\rm v}-\epsilon_\mu}
\label{valwf}
\ee
in the hadron tensor (\ref{stpnt}). The spinors 
$\psi_\mu(\bbox{r})=\langle\bbox{r}|\mu\rangle$
diagonalize the classical Dirac Hamiltonian (\ref{hamil}) yielding
the eigenvalues $\epsilon_\mu$. It is important to stress that the
dependence of the valence quark wave--function $\Psi_{\rm v}(\bbox{r},t)$
on the collective coordinates is included in the calculation of $g_1$. 
Only with this input it will be possible to disentangle the strange quark 
contribution. The calculation is performed by taking the Fourier 
transform of $\Psi_{\rm v}(\bbox{r},t)$ which allows us to carry out
the momentum integrals in eq (\ref{stpnt}). The technical details of 
this calculation will be presented elsewhere together with results
for the unpolarized structure functions \cite{Sch98}.

The matrix element in eq (\ref{stpnt}) between nucleon states is to be 
taken in the space of the collective coordinates, $A(t)$ 
(see eq. (\ref{colcor})) as the object in curly brackets is an operator 
in this space, which can be deduced from eq (\ref{valwf}). Here it is 
important to repeat that $A(t)$ spans the three dimensional flavor space 
which brings the strange degrees of freedom into the picture. This is the 
main difference to {\it e.g.} bag model calculations where the nucleon state 
is considered to be a product state of three specified quarks. In the 
present model it is rather a collective excitation of quark fields in 
the background of the self--consistent soliton.

\bigskip
\leftline{\large\it 5. Projection and Evolution}
\medskip

In the chiral soliton model the
baryons states are not momentum eigenstates causing the structure 
functions not to vanish exactly for $x>1$. This short--coming is due to 
the localized field configuration and thus the nucleon not being a 
representation of the Poincar{\'e} group. As a consequence the 
computed structure functions are not frame independent. It turns 
out that in the infinite momentum frame (IMF) the structure functions 
indeed vanish for $x>1$ \cite{Ja80,Ga97}. This is a consequence of the 
Lorentz contraction associated with the boost from the rest frame (RF) to
the IMF which exactly projects onto $0\le x\le 1$. For the polarized 
twist--2 structure function this implies
\be
g_1^{({\rm IMF})}(x)=\frac{1}{1-x}\, 
g_1^{({\rm RF})}\, \Big(-{\rm ln}(1-x)\Big)\ ,
\label{imf}
\ee
where $g_1^{({\rm RF})}(x)$ denotes the structure function computed 
in the nucleon rest frame as discussed in the preceding sections. The
transformation (\ref{imf}) leaves the integral over $g_1$ invariant
provided that in the rest frame the integration range has been extended to 
infinity.

The chiral soliton model is considered to approximate QCD at a low 
momentum scale $Q_0^2$ whence the result $g_1^{({\rm IMF})}(x)$
should be interpreted as $g_1(x,Q^2_0)$. It should be noted that 
$Q_0^2$ is a new parameter to the model. In order to compare with 
experimental data it is mandatory to evolve this structure function
to $g_1(x,Q^2)$ according to the DGLAP procedure \cite{Gr72}. 
Here $Q^2$ is the momentum scale set by the experiment. In order 
to apply this procedure we first have to separate the flavor singlet 
(0) and non--singlet ($ns$) contributions to $g_1(x,Q^2_0)$ as the 
former mixes with the gluon contribution $g(x,Q^2)$ upon evolution. This 
separation is straightforwardly accomplished by substituting the appropriate
flavor matrix for ${\cal Q}^2$ in eq (\ref{stpnt}). The leading 
order evolution equations read in the case of three flavors and with 
$t={\rm ln}(Q^2/\Lambda_{\rm QCD}^2)$ as well as
$\alpha_{\rm QCD}=4\pi/9t$,
\be
\frac{d g_1^{\rm (ns)}(x,t)}{dt}&=&
\frac{\alpha_{\rm QCD}}{2\pi}\int_x^1 \frac{dy}{y} 
P_{qq}\left(\frac{x}{y}\right) g_1^{\rm (ns)}(y,t)\ ,
\label{g1ns} \\
\frac{d g_1^{\rm (0)}(x,t)}{dt}&=&
\frac{\alpha_{\rm QCD}}{2\pi}\int_x^1 \frac{dy}{y} \left\{
P_{qq}\left(\frac{x}{y}\right) g_1^{\rm (0)}(y,t)
+6P_{qg}\left(\frac{x}{y}\right) g(y,t) \right\}\ ,
\label{g1s} \\
\frac{d g(x,t)}{dt}&=&
\frac{\alpha_{\rm QCD}}{2\pi}\int_x^1 \frac{dy}{y} \left\{
P_{gg}\left(\frac{x}{y}\right) g(y,t)
+P_{gq}\left(\frac{x}{y}\right) g_1^{\rm (0)}(y,t) \right\}\ .
\label{ggs}
\ee
The splitting functions $P_{ij}$ are listed in ref \cite{Al94}. Here it 
suffices to note that the integrals $\int_0^1 dz P_{qq}(z)$ 
and $\int_0^1 dz P_{qg}(z)$ vanish at leading order. As a result the 
zeroth moments 
\be
\triangle q = 2 \int_0^1 dx\, g_1^{(q)}(x,t)
\qquad {\rm for} \quad q=u,d,s
\label{zerothm}
\ee
do not depend on the momentum scale.
For the evolution program being applicable we have to 
assume that $g(x,Q_0^2)=0$ which implies that there a no 
soft gluons at the model scale while the effects attributed to
hard gluons are approximated by the contact interaction in (\ref{NJL}).
After integrating the differential equations (\ref{g1ns})--(\ref{ggs}) 
from $Q_0^2$ to $Q^2$ the singlet and non--singlet flavor components
are superposed to yield the desired flavor combination.

\bigskip
\leftline{\large\it 6. Numerical Results}
\medskip

In a first step we determine the model scale $Q_0^2$. For this purpose 
we perform the evolution for the unpolarized structure functions which 
enter the Gottfried sum rule. As this is a non--singlet combination 
a simplified evolution equation like (\ref{g1ns}) applies because gluon 
degrees of freedom do not contribute. We vary the lower boundary 
$Q_0^2$ when integrating the evolution equations until maximal agreement 
with the experimental data available at the upper boundary 
$Q^2=5{\rm GeV}^2$ is obtained. Since 
details of that calculation will be presented elsewhere \cite{Sch98} we 
content on quoting the result $Q_0^2=0.4{\rm GeV}^2$. This is identical 
to the value found for the two flavor model \cite{We96a}. Na\"{\i}vely 
one would expect that the two and three flavor models would give the same 
result because the strange degree of freedom cancels in this particular 
combination of structure functions. However, in the two and three flavor 
models the quark wave--functions are different as can easily be seen from 
eq (\ref{valwf}); in the considerably simpler two flavor version the 
only contributing perturbation is $\sum_{a=1}^3(\lambda_a/2)\Omega^a$,
to be contrasted with eqs (\ref{frf1})--({\ref{frf3}).

All presented results for the polarized structure function at the low 
momentum scale will be taken in the infinite momentum frame (\ref{imf}).
In table \ref{tab_1} we show the zeroth moments (\ref{zerothm}) of the 
polarized structure functions.
\begin{table}[tb]
\caption{\label{tab_1}
Zeroth moments of the polarized structure function $g_1$. Also given is 
the gluon component at $Q^2=3.0{\rm GeV}^2$, 
$\triangle G=2\int_0^1 dx g(x)$.}
\vspace{0.2cm}
\centerline{
\renewcommand{\arraystretch}{1.2}
\begin{tabular}{c|c|c|c|c}
$M_u({\rm MeV})$ & $\triangle u $ & $\triangle d $ 
& $\triangle s $ & $\triangle G $ \\
\hline
$400$ & $0.64$ & $-0.14$ & $-0.01$ & $0.23$ \\
$450$ & $0.60$ & $-0.16$ & $-0.02$ & $0.21$
\end{tabular}}
\renewcommand{\arraystretch}{1.0}
\end{table}
In ref \cite{Bl96} the vacuum contributions to these moments were computed. 
For the sum of valence and vacuum parts those authors give\footnote{In 
the present calculation PCAC violating $1/N_C$ contributions to the 
non--singlet combinations have been ignored. Hence we have to compare our 
results with the entry ``NJL(scalar)'' of table X in ref \cite{Bl96}.
In ref \cite{Bl96} a different regularization scheme to determine the
chiral angle and a different expansion scheme of the fermion determinant 
were used. This might as well make the (small) differences in the zeroth 
moments.} $\triangle u=0.64$, $\triangle d=-0.24$ and $\triangle s=-0.02$ 
using $M_u=423{\rm MeV}$. Apparently the vacuum contribution to the zeroth 
moment is small in accord with our valence quark approximation to the 
polarized structure functions. We also note that due to the deviation 
from SU(3) symmetric baryon wave--functions the quantities shown in table 
\ref{tab_1} should not be related to the $F$ and $D$ parameters determined 
from semi--leptonic hyperon decays\footnote{Strictly speaking $F$ and $D$ 
parameters are only well--defined in a flavor symmetric formulation.}.

In figure \ref{fig_1} the main result of our calculation is displayed:
the strangeness contribution $g_1^{(s)}$ to the polarized nucleon 
structure function $g_1$.
\begin{figure}[ht]
~
\vskip1cm
\centerline{\hskip 0.5cm
\epsfig{figure=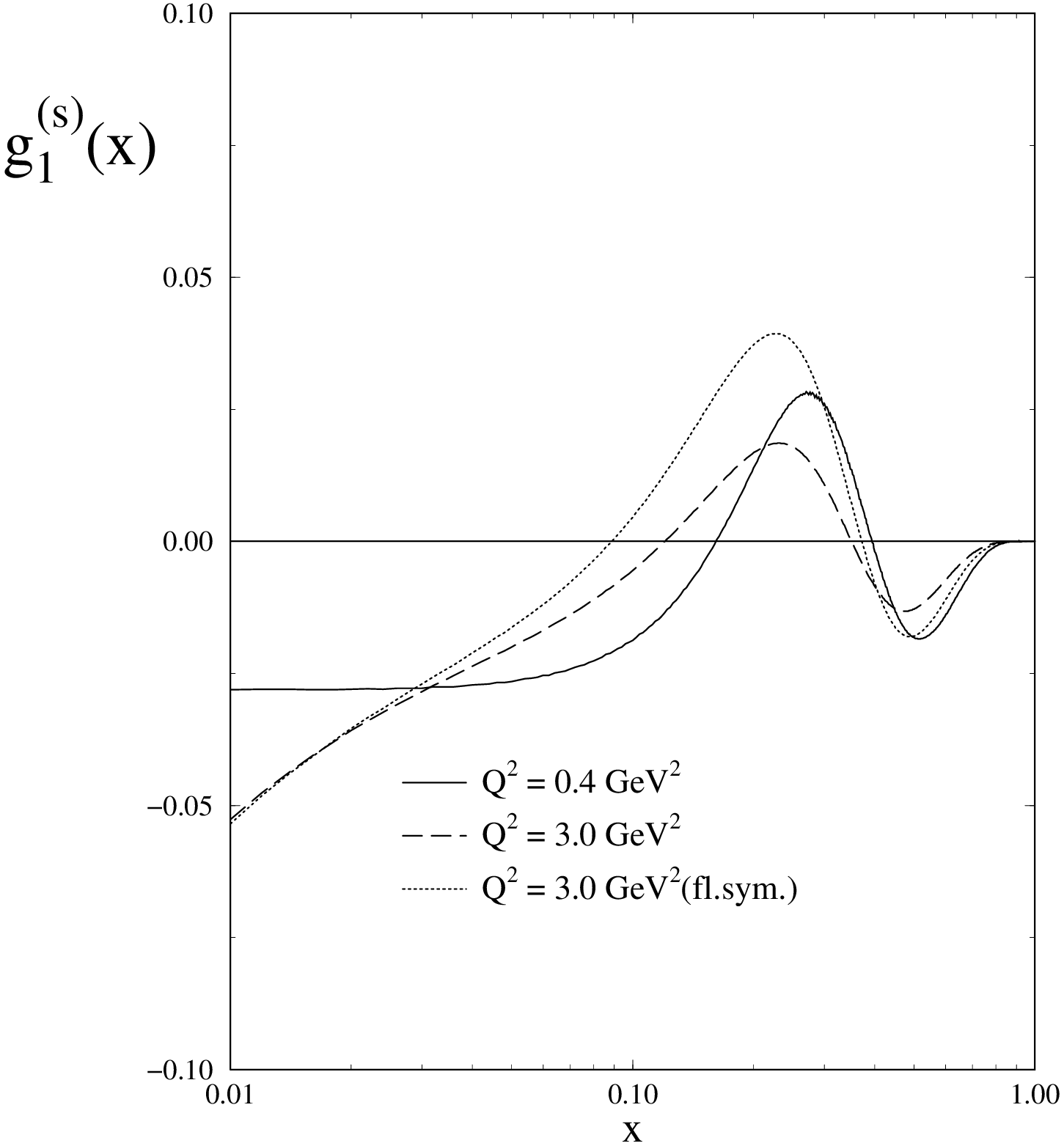,height=6cm,width=7.5cm}
\hskip 1.0cm
\epsfig{figure=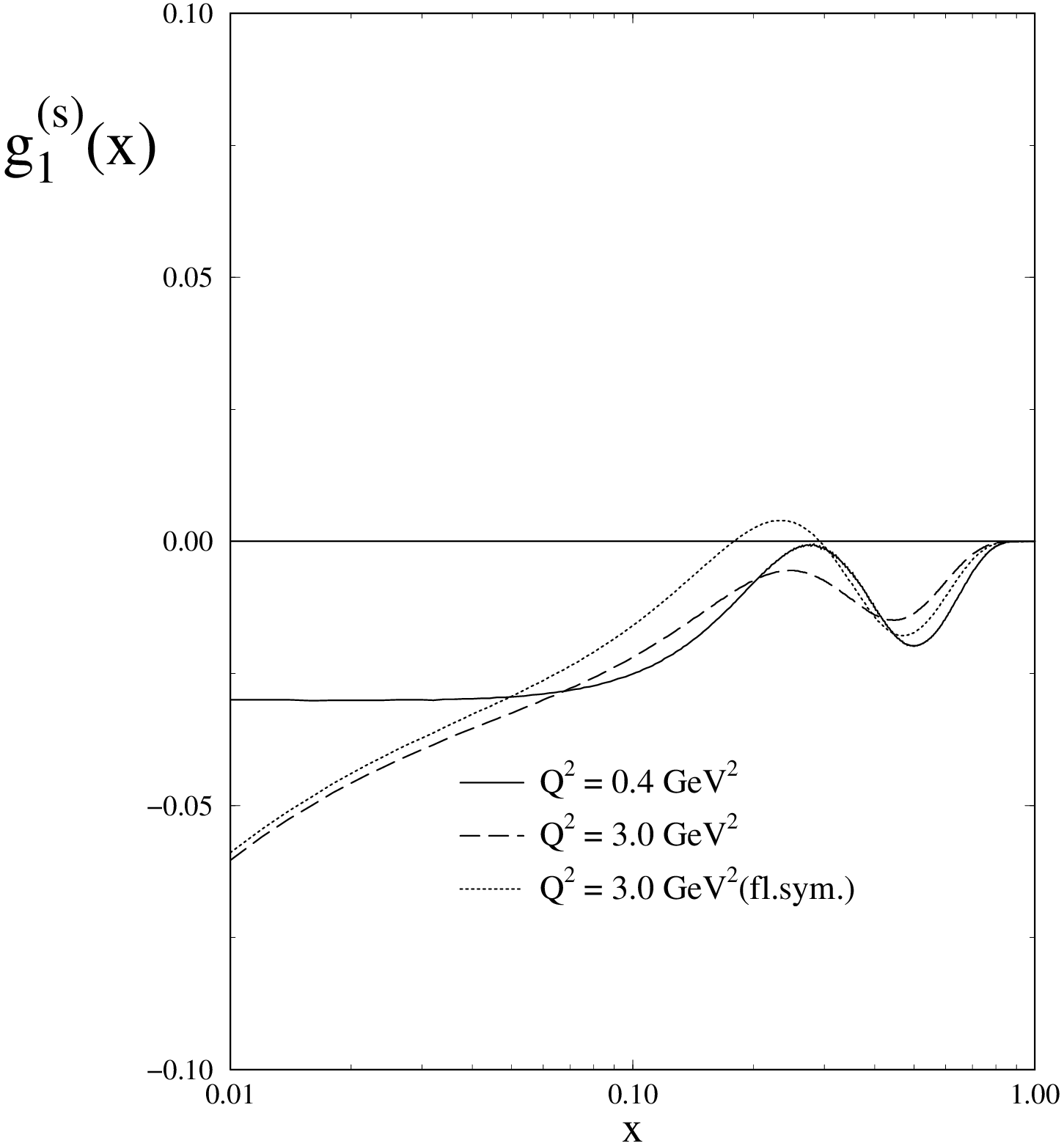,height=6cm,width=7.5cm}}
\vskip0.5cm
\caption{\label{fig_1}
The strange quark contribution to the polarized structure function 
$g_1$ for the proton in the infinite momentum frame at the model scale 
$Q^2=0.4{\rm GeV}^2$ (full line). Two cases are displayed, 
$M_u=400{\rm MeV}$ (left panel) and $M_u=450{\rm MeV}$ (right panel). 
Also shown is the leading order QCD evolution to $Q^2=3{\rm GeV}^2$
(dashed line). Furthermore the strange quark contribution computed
with a pure octet nucleon wave--function (after evolution) is 
displayed (dotted line).}
\end{figure}
Apparently the smallness of $\triangle s$ does not necessarily transfer 
to an overall negligible $g_1^{(s)}$. A cancellation between positive 
and negative parts is not excluded. This effect is not altered by 
the DGLAP evolution. We furthermore compare to the strange quark 
contribution obtained with a pure octet nucleon wave--function, {\it i.e.}
with the symmetry breaking effects omitted when diagonalizing (\ref{collham}).
Apparently the incorporation of symmetry breaking effects in the nucleon 
wave--function via exact diagonalization of the collective Hamiltonian 
yields significantly less pronounced strange quark contributions. 
This is not unexpected in the collective approach \cite{We96}.

In figure \ref{fig_2} we compare the two and three flavor model predictions
for the electromagnetic flavor combination ${\cal Q}^2$ for which data
from SLAC are available at $Q^2=3GeV^2$ \cite{Abe95}.
\begin{figure}[ht]
~
\vskip1cm
\centerline{\hskip 0.5cm
\epsfig{figure=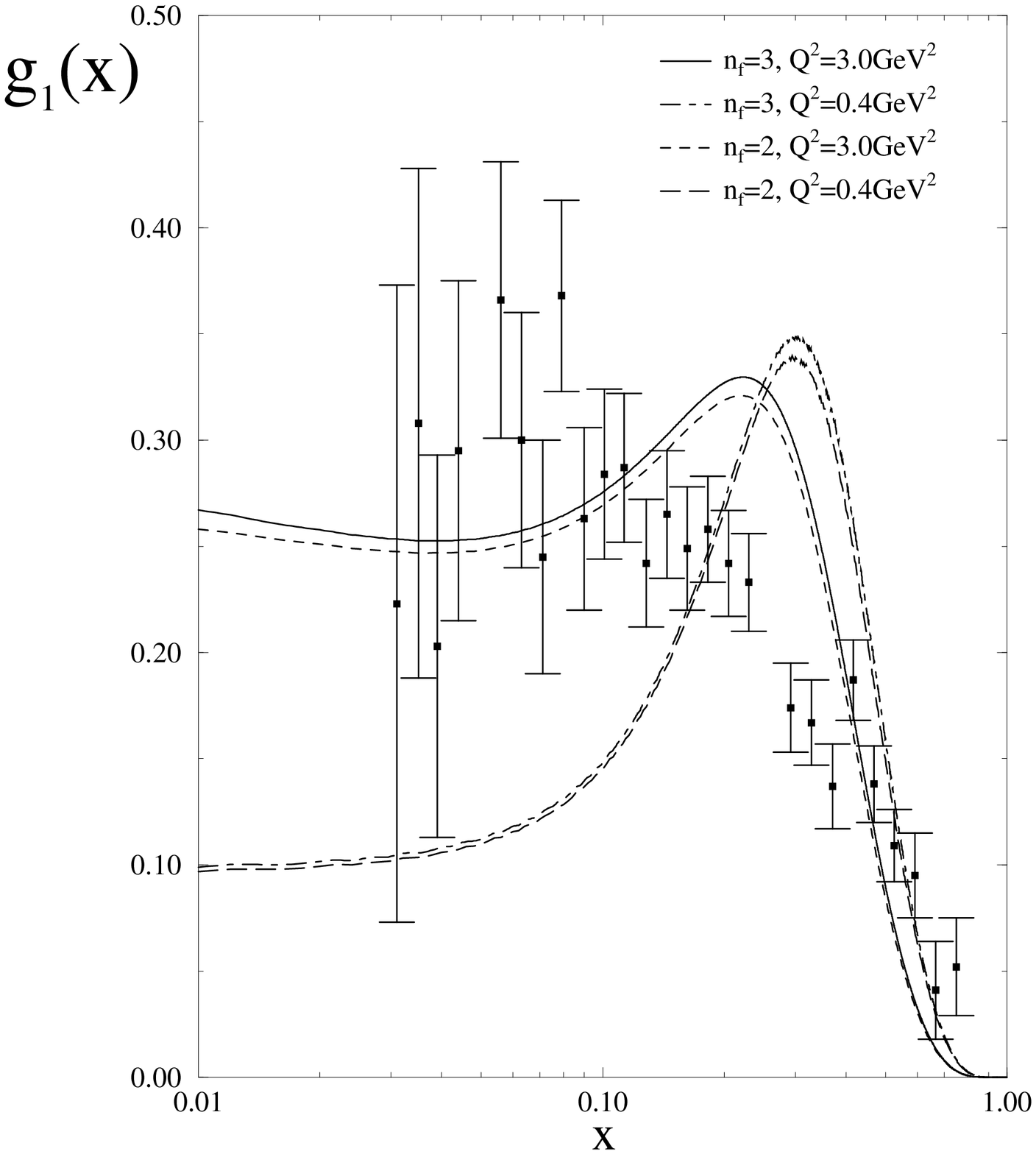,height=6cm,width=7.5cm}
\hskip 1.0cm
\epsfig{figure=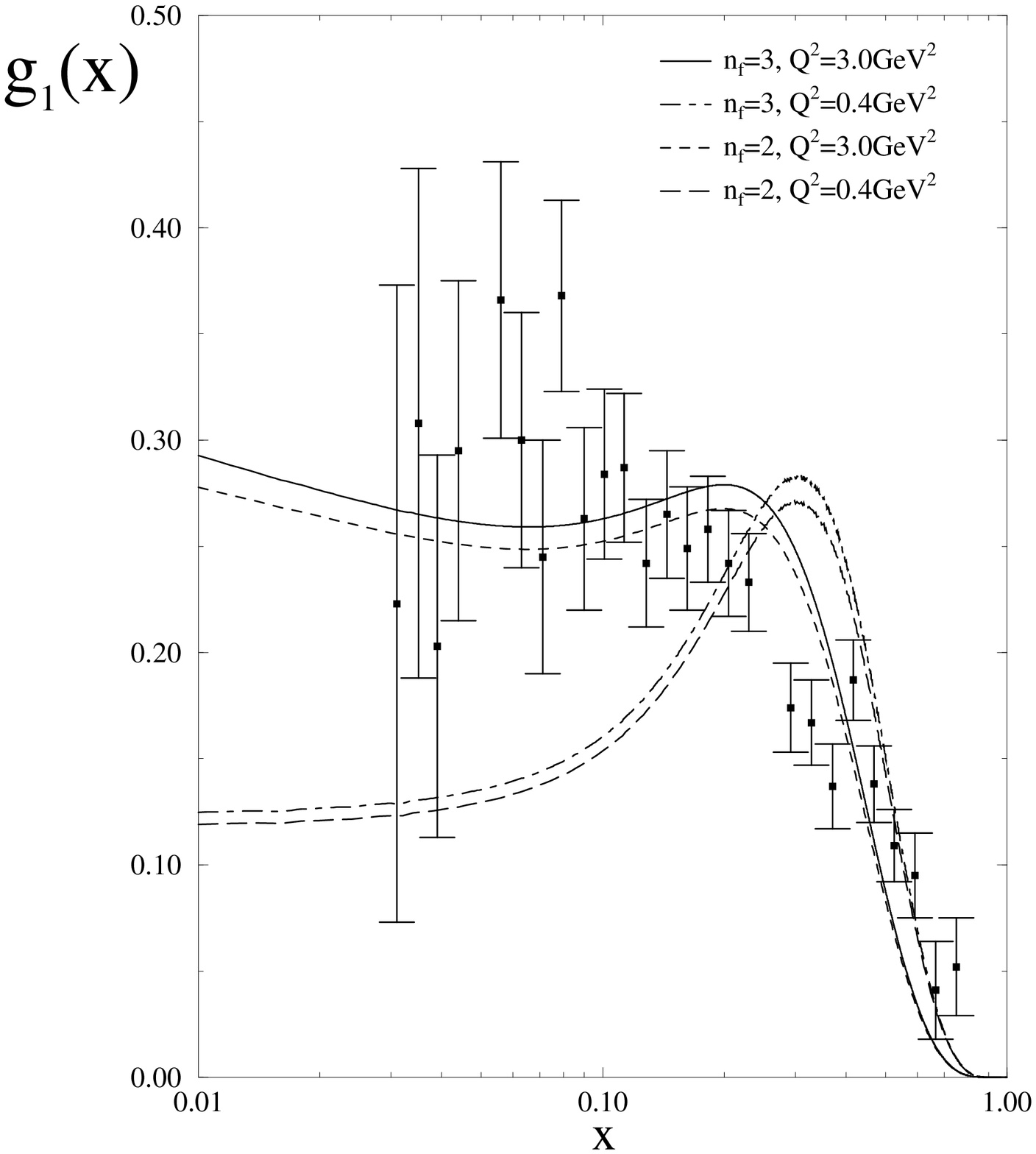,height=6cm,width=7.5cm}}
\vskip0.5cm
\caption{\label{fig_2}
Comparison of the two and three flavor calculation of the 
polarized structure function $g_1$ for the proton in the 
infinite momentum frame. Also shown is the leading order 
QCD evolution to $Q^2=3{\rm GeV}^2$. Left panel: $M_u=400{\rm GeV}$,
right panel: $M_u=450{\rm GeV}$. Data are taken from ref
\protect\cite{Abe95}.}
\end{figure}
We recognize that the inclusion of the strangeness degree of freedom
yields only minor changes. For $M_u=400{\rm MeV}$ the results obtained
in the two and three flavor models are almost indistinguishable.
In particular both versions reasonably reproduce the SLAC data 
\cite{Abe95}. The small change of $g_1(x)$ when generalizing to 
flavor SU(3) is unexpected, after all there are at least two 
significant effects associated with this generalization. First,
strange quarks appear explicitly, second in SU(3) the Clebsch--Gordan
coefficients are different and so are the nucleon matrix elements 
of the collective operators\footnote{The inclusion of symmetry
breaking effects within the Yabu--Ando approach drives these matrix 
elements towards their SU(2) values.}. Apparently these two 
effects partially cancel each other.

In figure \ref{fig_3} we show our prediction for the gluon component
of the polarized structure function at the experimental scale 
$Q^2=3{\rm GeV}^2$. We remind the reader of the assumption that at 
the model scale ($Q_0^2=0.4{\rm GeV}^2$) this component is taken 
to be zero, {\it i.e.} this component is solely due to radiation and 
absorption of soft gluons. We remark that the singularity at 
$x\to0$ is weaker than $1/x$ since $xg(x)\to0$ in this limit.
\begin{figure}[ht]
~
\vskip1cm
\centerline{\hskip -1.0cm
\epsfig{figure=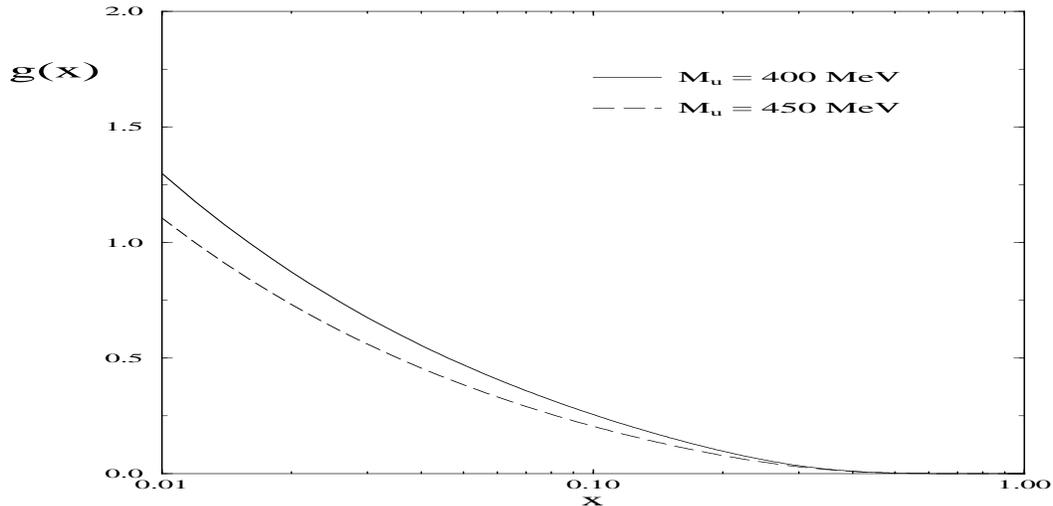,height=6cm,width=14cm}}\vskip1cm
\caption{\label{fig_3}
The gluon component of the polarized structure function evolved from 
the model scale $Q_0^2$ to $Q^2=3{\rm GeV}^2$. The results are displayed 
for two different values of the up constituent quark mass $M_u$.}
\end{figure}

\bigskip
\leftline{\large\it 7. Conclusions}

In this letter we have presented some first results for strange quark 
contributions to nucleon structure functions. In particular we have 
concentrated on the twist--2 piece of the polarized structure function 
$g_1$. On the whole we find the surprising result that the contribution of 
the strange degrees of freedom do not significantly alter the results of 
the pure two flavor model. In both the two and three flavor versions 
of the NJL chiral soliton model we obtain reasonable agreement with 
experimental data for the polarized nucleon structure function $g_1$ when 
the scale dependence is accounted for according to the DGLAP scheme.

To calculate various flavor components of $g_1$ at the low momentum 
scale we have employed the valence quark approximation to the NJL
chiral soliton model in flavor SU(3). Although the valence quark
contribution is known to strongly dominate the static axial properties
for the self--consistent chiral soliton, in the unpolarized case some 
conceptual requirements like positivity are slightly violated by this 
approximation \cite{We96a}. Hence the full calculation would require to 
include the effects attributed to the polarized vacuum as well. As 
discussed, we do not expect them to have significant impact on the 
result. In addition an extensive next--to--leading order DGLAP 
evolution could be performed for the twist--2 structure functions
under consideration. As the main effect we would expect a variation
of the model scale $Q_0^2$, which anyhow is a free parameter.

\end{document}